\documentclass[12pt]{article}

\usepackage{epsfig}

\def\beq{\begin{equation}}
\def\eeq{\end{equation}}
\def\bea{\begin{eqnarray}}
\def\eea{\end{eqnarray}}

\def\gev{\, {\rm GeV}}
\def\mev{\, {\rm MeV}}

\newcommand{\gsim}{\lower.7ex\hbox{$\;\stackrel{\textstyle>}{\sim}\;$}}
\newcommand{\lsim}{\lower.7ex\hbox{$\;\stackrel{\textstyle<}{\sim}\;$}}

\begin{document}

\setlength{\baselineskip}{0.25in}


\begin{titlepage}
\noindent
\begin{flushright}
MCTP-03-44 / UT-03-33\\
\end{flushright}
\vspace{1cm}

\begin{center}
  \begin{Large}
    \begin{bf}
Neutrino-induced lepton flavor violation \\
in gauge-mediated supersymmetry breaking
    \end{bf}
  \end{Large}
\end{center}
\vspace{0.2cm}
\begin{center}
\begin{large}
Kazuhiro Tobe$^{a,b}$, James D. Wells$^{a}$, Tsutomu Yanagida$^{c}$ \\
\end{large}
  \vspace{0.3cm}
  \begin{it}
${}^{(a)}$Michigan Center for Theoretical Physics (MCTP) \\
        ~~University of Michigan, Ann Arbor, MI 48109-1120, USA \\
\vspace{0.1cm}
${}^{(b)}$Department of Physics, University of California, Davis 95616, USA \\ 
\vspace{0.1cm}
${}^{(c)}$Department of Physics, University of Tokyo, Tokyo 113-0033, Japan
  \end{it}

\end{center}

\begin{abstract}

Gauge-mediated supersymmetry breaking is known to greatly suppress
flavor changing neutral current effects.  However, we show that
gauge mediation in the context of leptogenesis implies potentially
large lepton flavor violating signals.  If the heavy right-handed neutrinos
that participate in leptogenesis are lighter than the messenger scale of
gauge mediation, they will induce flavor off-diagonal masses to the
sleptons which in turn can induce large effects in $\mu\to e\gamma$,
$\tau\to \mu\gamma$, and $\mu -e$ conversion in nuclei.
We demonstrate this result and compute numerically the
lepton-flavor violating decay and conversion rates
in scenarios of direct gauge mediation.

\end{abstract}

\vspace{1cm}

\begin{flushleft}
hep-ph/0310148 \\
October 2003
\end{flushleft}

\end{titlepage}


\section{Introduction}

One of the challenges of supersymmetry is constructing a theory
that has no new sources of flavor violation substantially beyond
those already present in the Standard Model (SM). This requirement
is thrust upon us by experiment, yet there are theory ideas that
predict it. 

One such flavor-tame theory is gauge mediated supersymmetry
breaking (GMSB).
In minimal models~\cite{Dine:1994vc,Giudice:1998bp} of GMSB (mGMSB)
there are $n_m$ copies of $5+\bar 5$ messengers that feel supersymmetry
breaking via a gauge singlet, $S=M+\theta^2 F_S$, in which $M$ is a
messenger mass and $F_S$ represents a supersymmetry breaking scale 
in the messenger sector. Since the messenger particles
interact with minimal supersymmetric SM (MSSM) particles only through
the SM gauge forces,  the messenger-loop diagrams give rise to 
flavor-diagonal supersymmetry breaking masses in the MSSM sector at the
messenger scale $M$, and hence flavor violation in supersymmetry breaking
masses are automatically suppressed at the messenger scale 
$M$~\cite{Dine:1994vc}.

The messenger scale can be any numerical value in principle.  From a collider
physics standpoint it is very interesting when $M$ and $F$ (goldstino
auxiliary component) are as low as possible as it can potentially lead
to prompt decays of 
the lightest SM superpartner into its corresponding SM particle plus
gravitino~\cite{Giudice:1998bp}. 
The collider signatures for this are usually easy
to distinguish from background and if found would provide a compelling
case for gauge mediation or some close variant of the theory.

Despite the interesting collider phenomenology of low-scale messengers,
there are some advantages to considering a much higher messenger
scale. This is particularly true when we consider how massive neutrinos
fit into the full theory, as nowadays results from
neutrino oscillation experiments clearly require.  
These neutrinos are best described by a seesaw mechanism with very heavy
right-handed neutrinos, $N_i$~\cite{Seesaw}. 
Out of equilibrium decays of these heavy neutrinos can precipitate
leptogenesis, leading to proper baryogenesis through sphaleron
reprocessing~\cite{Fukugita:1986hr}. 
Applying all these criteria we expect the right-handed neutrino masses to
lie somewhere in the range~\cite{Asaka:1999yd,Hamaguchi:2001gw}
\bea
10^{6-8}\gev \lsim M_{R_i}\lsim 10^{12-16}\gev .
\eea

Making leptogenesis work requires that the reheat temperature be
at least as large as $T_R>10^6\gev$~\cite{Asaka:1999yd,Hamaguchi:2001gw}. 
Gravitinos can be produced
copiously during the reheat phase of the universe~\cite{Moroi:1995fs}.  
For any given
reheat temperature there is a lower bound on the mass of the gravitino
such that the universe does not become overclosed. Near this overclosure
limit, the gravitino can be a significant component of the dark matter.
For $T_R=10^6\gev$
that lower limit is $m_{3/2}>100\mev$.\footnote{
If there is an entropy production after the freeze-out time of the
gravitino, this limit would not be applicable because it dilutes the
number density of the gravitino~\cite{Fujii:2002fv}.}
Thus, supergravity tells
us
\beq
\label{gravitino mass}
m_{3/2}=\frac{F}{\sqrt{3}M_{\rm Pl}}\gsim 100\mev 
\Longrightarrow \sqrt{F}\gsim {\rm few}\, \times 10^8\gev .
\eeq
Keeping flavor-mixing supergravity contributions to scalar masses
sufficiently suppressed is equivalent to requiring $m_{3/2}\lsim 10\gev$.
Therefore, we have a lower limit and an upper limit for $F$,
\beq
\label{allowed F}
10^{17}\gev^2 \lsim F \lsim 10^{19}\gev^2 ~~{\rm
(allowed~range~for~}F{\rm ).} 
\label{F_condition}
\eeq

The supersymmetry breaking $F_S$-term(s) that participate in GMSB
are not necessarily the full strength $F$-term(s) that add to the
gravitino mass.  In the early Dine-Nelson GMSB models~\cite{Dine:1994vc}, 
$F_S$ was a loop factor below the $F$ in 
Eq.~(\ref{gravitino mass}). However, there is a class of GMSB models,
which we call direct gauge mediation, where
$F_S=F$~\cite{Izawa:1996pk,Intriligator:1996pu,Arkani-Hamed:1997jv,
Dimopoulos:1997ww,Murayama:1997pb,Izawa:1997gs}.
Combining the collider limits on the gaugino masses
$M_i\gsim 100\gev$ with the  allowed range for $F$ obtained from
requirements on the gravitino mass in direct gauge mediation, 
we can estimate the allowed range for the messenger mass, 
\beq
10^{12}\gev \lsim M \lsim 10^{15}\gev ~~~ ({\rm messenger~mass~range}).
\label{M_condition}
\eeq
Interestingly successful leptogenesis induced by
heavy right-handed neutrinos leads to high messenger scale $M$ in 
direct gauge mediation. 

Under the requirements in Eqs.~(\ref{F_condition}, \ref{M_condition}),
the phenomenology no longer admits prompt decays of the lightest
superpartner into a gravitino. If the lightest SM superpartner is a
neutralino the phenomenology will be hard to distinguish from a regular
supergravity model.\footnote{A charged lightest 
SM superpartner that is long-lived on collider time scales would be
assumed to be metastable since copious charged dark matter is not
allowed cosmologically.  Therefore, one would conclude that it must
decay to something else, and a light gravitino would be the leading
candidate for a decay product that carries away odd R-parity.}
However, the interesting phenomenology of these models does not stop
with collider physics.  Indeed, in GMSB with such a high-messenger
scale,  heavy right-handed neutrinos play an important role
in low-energy phenomenology as well as in leptogenesis.
Perhaps the most important and unique phenomenological implication of
this theory is the potentially large lepton flavor violating (LFV) 
signals that can arise. 
This is counter-intuitive since GMSB is
usually thought to not allow superpartner-induced LFV amplitudes.  In
this article  we demonstrate why LFV happens in this general class of
supersymmetry theories, and compute the expected rates
for LFV observables $B(\mu\to e\gamma)$, $B(\tau\to \mu\gamma)$,
and $\mu-e$ conversion in nuclei.\footnote{Within the framework of
supergravity supersymmetry breaking, leptogenesis and LFV have been
discussed in Ref.~\cite{Ellis:2002xg}.}

\section{Supersymmetric SM with right-handed neutrinos}

We begin by reviewing the MSSM with right-handed neutrinos, in which
the seesaw mechanism naturally induces tiny neutrino masses.
The superpotential in the lepton sector is given by
\begin{eqnarray}
W=y_e^i E_i L_i H_1+y_\nu^{i,j} N_i L_j H_2 +\frac{M_{R_i}}{2}N_i N_i.
\end{eqnarray}
Here we take a basis where the charged lepton Yukawa matrix $y_e$ and 
the right-handed neutrino mass matrix $M_R$ are both diagonal.
Note that in general the neutrino Yukawa couplings $y_\nu$ 
are not diagonal and become a source for LFV
phenomena such as neutrino oscillation, $\mu \rightarrow e \gamma$, etc.

Below each right-handed neutrino mass scale $M_{R_i}$, the corresponding
right-handed neutrino $N_i$ decouples. We write the resulting low-energy 
effective
superpotential for the neutrino masses at the electroweak scale
as follows,
\begin{eqnarray}
W_{\rm eff} &=& -\frac{\kappa_{ij}}{2} (L_i H_2) (L_j H_2),
\\
\kappa_{ij}&\equiv& \left( y_\nu^{\rm T} M_R^{-1} y_\nu
\right)_{ij}.
\label{kappa_matrix}
\end{eqnarray}
The neutrino mass matrix then is given by
\begin{eqnarray}
m_\nu = \kappa \langle H_2^0 \rangle^2
=\frac{\kappa v^2 \sin^2 \beta}{2},
\end{eqnarray}
which can be diagonalized by the Maki-Nakagawa-Sakata (MNS) 
matrix $U^{\rm MNS}$~\cite{Maki:mu}:
\begin{eqnarray}
(U^{\rm MNS})^{\rm T} m_\nu U^{\rm MNS}={\rm diag}
(m_{\nu_1},m_{\nu_2},m_{\nu_3}),
\\
(U^{\rm MNS})^{\rm T} \kappa U^{\rm MNS}={\rm diag}
(\kappa^D_1,\kappa^D_2,\kappa^D_3).
\label{kappa}
\end{eqnarray}
From Eqs.~(\ref{kappa_matrix}) and (\ref{kappa}), the neutrino Yukawa matrix 
can be expressed as~\cite{Casas:2001sr},
\begin{eqnarray}
y_\nu^{ij} = \sqrt{M_{R_i}} R_{ik} \sqrt{\kappa_k^D}
(U^{\rm MNS *})_{jk}, 
\label{neutrino_yukawa}
\end{eqnarray}
where the matrix $R$ is an unknown orthogonal matrix $R^{\rm T} R=${\bf 1}.
Low-energy neutrino oscillation experiments determine the neutrino 
mass differences and the entries of the MNS matrix. Here we parameterize 
the MNS matrix as follows:
\begin{eqnarray}
U^{\rm MNS} =
\left(
\begin{tabular}{ccc}
$c_{12} c_{13}$ & $s_{12} c_{13}$ & $s_{13} e^{-i \delta}$\\
$-s_{12} c_{23} -c_{12} s_{23} s_{13} e^{i \delta}$ & 
$c_{12} c_{23}-s_{12} s_{23} s_{13} e^{i\delta}$ & $s_{23} c_{13}
e^{i\delta}$ \\
$s_{12} s_{23}-c_{12} c_{23} s_{13} e^{i \delta}$ & 
$-c_{12} s_{23}-s_{12}c_{23} s_{12} e^{i \delta}$ &$c_{23} c_{13}$
\end{tabular}
\right),
\end{eqnarray}
where $s_{ij}=\sin\theta_{ij}$ and $c_{ij}=\cos\theta_{ij}$.

In our analysis, we adopt the following numerical 
data~\cite{ATMO,Ahmed:2003kj}:
\begin{eqnarray}
\sin^2 2\theta_{23}&=&1,\\
\Delta m^2_{\rm atm.}&=&m^2_{\nu_3}-m^2_{\nu_2} =
2\times 10^{-3}~{\rm eV}^2,
\end{eqnarray}
for atmospheric neutrinos, and 
\begin{eqnarray}
\theta_{12} &=&32.5~{\rm degrees},\\
\Delta m^2_{\rm solar} &=& m^2_{\nu_2}-m^2_{\nu_1}=
7.1\times 10^{-5}~{\rm eV}^2,
\end{eqnarray}
for solar neutrinos. We assume hierarchical neutrino
masses, so that we take $m_{\nu_1}=0$ in our analysis.
Angle $\theta_{13}$ has not been determined yet, and its experimental
limit is $|s_{13}|\leq 0.2$~\cite{ATMO}. In our numerical analysis, we
will take $s_{13}=0$ or $0.1$ as reference values.
For simplicity, we neglect complex phase $\delta$.
\section{Lepton flavor violation in GMSB}

Off-diagonal components in the neutrino Yukawa couplings give rise to
LFV neutrino oscillations, and they also can induce significant
LFV in the charged lepton sector.
For example, even if the supersymmetry breaking mechanism
is flavor-blind, as is the case with GMSB, 
LFV in the slepton masses can be induced through radiative
corrections (i.e., renormalization group running)
involving neutrino Yukawa
interactions~\cite{Borzumati:1986qx,Hisano:1995nq}. However, effects of
neutrino Yukawa couplings decouple 
if the scale of supersymmetry breaking is
lower than the right-handed neutrino mass scales. In that case, the LFV in
the slepton mass matrix is negligibly small. 
Low-energy GMSB behaves in this manner if the messenger scale is 
less than the right-handed neutrino masses.

However, in the case of high-scale direct gauge mediation, which we
motivated in the introduction, 
the messenger scale
where supersymmetry breaking terms are induced can be higher than the
right-handed neutrino mass scales, $M>M_{R_i}$. In this
case, the neutrino Yukawa couplings enter the slepton mass matrix
renormalization group equation over a significant scale range, thereby
affecting potentially large LFV in the charged lepton sector.\footnote{
Since the messenger scale $M$ is lower than the GUT scale,
GUT-induced LFV in superpartner masses is absent.}

We can see this LFV effect through the renormalization group equations
(RGEs) for the left-handed slepton masses:
\begin{eqnarray}
\mu \frac{d}{d\mu} (m^2_{\tilde{L}})_{ij} &=& 
\mu \frac{d}{d\mu} (m^2_{\tilde{L}})_{ij}|_{\rm MSSM}
\nonumber \\ &+&
\frac{1}{16\pi^2} \left[
m^2_{\tilde{L}} y^\dagger_\nu y_\nu +y^\dagger_\nu y_\nu 
m^2_{\tilde{L}} +2 \left(y^\dagger_\nu m^2_{\tilde{\nu}} y_\nu
+m^2_{H_2} y^\dagger_\nu y_\nu +A^\dagger_\nu A_\nu\right)
\right]_{ij},
\label{RGE_slepton}
\end{eqnarray} 
where $\mu d(m^2_{\tilde{L}})_{ij}/d \mu|_{\rm MSSM}$ is a 
contribution from MSSM without right-handed neutrinos. 
Even if $m^2_{\tilde{L}}$ and $m^2_{\tilde{\nu}}$
are flavor-diagonal at the messenger scale, flavor mixings in the neutrino
Yukawa couplings generate flavor violation in the left-handed slepton masses 
$m^2_{\tilde{L}}$. The flavor mixing in the left-handed slepton masses
is approximately proportional to $y^\dagger_\nu y_\nu$.
Using Eq.~(\ref{neutrino_yukawa}),
\begin{eqnarray}
(y^\dagger_\nu y_\nu)_{\mu e} &=&
(U^{\rm MNS} \sqrt{\kappa^D} R^\dagger M_{R} R \sqrt{\kappa^D}
U^{\rm MNS~\dagger})_{\mu e}, \\
&\simeq& U^{\rm MNS}_{\mu 3} U^{\rm MNS *}_{e3}
\kappa^D_3 M_{R_3} \left(|R_{33}|^2 
+\frac{M_{R_2}}{M_{R_3}} |R_{23}|^2
+\frac{M_{R_1}}{M_{R_3}} |R_{13}|^2 \right) \nonumber \\
&&+ U^{\rm MNS}_{\mu 3} U^{\rm MNS *}_{e2}
\sqrt{\kappa^D_3 \kappa^D_2} M_{R_3} \left(
R_{33}^* R_{32} +\frac{M_{R_2}}{M_{R_3}} R_{23}^* R_{22}
+\frac{M_{R_1}}{M_{R_3}} R_{13}^* R_{12} \right) \nonumber \\
&&+ U^{\rm MNS}_{\mu 2} U^{\rm MNS *}_{e 3}
\sqrt{\kappa^D_2 \kappa^D_3} M_{R_3} \left(
R_{32}^* R_{33} +\frac{M_{R_2}}{M_{R_3}} R_{22}^* R_{23}
+\frac{M_{R_1}}{M_{R_3}} R_{12}^* R_{13} \right) \nonumber \\
&&+ U^{\rm MNS}_{\mu 2} U^{\rm MNS *}_{e 2}
\kappa^D_2 M_{R_3} \left(
|R_{32}|^2 +\frac{M_{R_2}}{M_{R_3}} |R_{22}|^2
+\frac{M_{R_1}}{M_{R_3}} |R_{12}|^2 \right).
\label{yy}
\end{eqnarray} 
In Eq.~(\ref{yy}), we assumed $\kappa^D_1 \ll \kappa^D_{2,3}$.

For $\tau \rightarrow \mu$ flavor violation, indices $\mu$ and $e$
should be replaced by $\tau$ and $\mu$, respectively.
As can be seen from Eq.~(\ref{yy}), LFV is induced by flavor mixings
from $U^{\rm MNS}$ and $R$.
Although the matrix $R$ is not known, the large neutrino mixing angles
in $U^{\rm MNS}$ suggested by the atmospheric and solar
neutrino experiments imply that large LFV effects are possible when
at least one $M_{R_i}<M$.\footnote{
The hierarchical neutrino Yukawa couplings also induce non-degeneracy
in diagonal components of the left-handed slepton mass matrix.
Therefore, the precise measurement of slepton and sneutrino masses
also would be important to probe the effects of the neutrino Yukawa
couplings.}

In this article we mainly adopt mGMSB to calculate event rates of LFV
processes. In mGMSB, MSSM gaugino ($M_i$, $i=1-3$) and sfermion masses
are generated at the messenger scale $M$ as follows:
\bea
\label{mGMSB masses}
M_i (M)& = & n_m \frac{\alpha_i(M)}{4\pi}\frac{F_S}{M} \\
\tilde m^2(M) & = & \frac{n_m}{8\pi^2}
\left\{c_1\alpha_1^2(M) +c_2\alpha_2^2(M)+c_3\alpha_3^2(M)\right\}
\frac{F_S^2}{M^2}
\eea
where $\alpha_i$ are $SU(5)$-GUT normalized gauge couplings.
The $c_i$ are the Casimirs of the sparticle
representations under the SM gauge groups, which in the SM translates to 
\beq
c_1=\frac{3}{5}\left( \frac{Y}{2}\right)^2,~~~
c_2=\left\{ \begin{array}{c} 3/4,~{\rm if}~$SU(2)$~{\rm doublet} \\
                             0,~{\rm if}~$SU(2)$~{\rm singlet}
            \end{array}\right. ,~~~
c_3=\left\{ \begin{array}{c} 4/3,~{\rm if}~SU(3)~{\rm triplet} \\
                             0,~{\rm if}~SU(3)~{\rm singlet}
            \end{array}\right.
\label{casimirs}
\eeq
Below the messenger scale $M$, we consider the MSSM with massive right-handed
neutrino(s) as the effective theory. We then numerically compute RGEs from the
messenger scale $M$ to
electroweak scale, taking into account decoupling of the right-handed
neutrino $N_i$ at each mass scale $M_{R_i}$. During the RG running
between $M$ and $M_{R_i}$, neutrino Yukawa interactions generate LFV
in the slepton masses as shown in Eq.~(\ref{RGE_slepton}),
and as a result,
low-energy LFV events are induced.

In order to show the messenger-model-dependence of the result, we also
consider the $Q$-$S$-$\phi$ model of Ref.~\cite{Murayama:1997pb} as an
example that has a complicated messenger sector.\footnote{
We call the model of Ref.~\cite{Murayama:1997pb} the ``$Q$-$S$-$\phi$ model''
because it contains three messenger mass scales $M_Q$, $M_S$ and
$M_\phi$ associated with $Q$, $S$ and $\phi$ states, where
$M_Q=\lambda v_Q/\sqrt{5}~ (\sim 10^{15}~{\rm GeV})$,
$M_S=(v_Q/\sqrt{5})^3/M_{\rm Pl}^2~ (\sim 10^8~{\rm GeV})$ and
$M_\phi=(v_Q/\sqrt{5})^4/M_{\rm Pl}~ (\sim 10^5~{\rm GeV})$.} 
In this model, at the highest energies the
gauge group is $SU(5)_H\times SU(5)_G$ with gauge couplings $\alpha_H$
and $\alpha_G$ respectively.  We assume all SM matter fields couple only to 
$SU(5)_H$.  Below the threshold of the $Q$ states $M_Q$,
the gauge group breaks to $SU(5)_G\times SU(5)_H
\to SU(5)$ with resulting gauge coupling $\alpha$. The low-energy 
superpartner spectrum at the scale $M_Q$
can be computed using the techniques of Ref.~\cite{Giudice:1997ni}:
\bea
M_{i}(M_Q) & = & 
\frac{\alpha(M_Q)}{4\pi}\frac{F}{M_Q}
(b_S-b_H-b_G) \\
\tilde m^2(M_Q) & = &
\frac{c\, \alpha^2(M_Q)}{8\pi^2}\frac{F^2}{M_Q^2}
\left\{b_H\frac{\alpha_H^2(M_Q)}{\alpha^2(M_Q)}+
b_S-2(b_H+b_G)\right\}.
\label{murayama mass}
\eea
where $c$ is defined in Eq.~(\ref{casimirs}), and the $b_i$ are defined
to be the coefficients of the gauge $\beta$ functions (see Eq.~(5) and just
below Eq.~(84) of Ref.~\cite{Giudice:1997ni} for explanation of these
values). In our numerical analysis, we will assume
$\alpha_H^2(M_Q)/\alpha^2(M_Q)=1$ for simplicity.\footnote{In the case
with $\alpha_H^2(M_Q)/\alpha^2(M_Q)=1$, $A$-terms vanish at $M_Q$.}
It is necessary to apply the above scalar mass equation for 
each gauge group the superpartner scalar is charged under, and
then sum the result to obtain the mass. At thresholds $M_S$ and $M_\phi$
where chiral messenger fields $S$ and $\phi$ decouple respectively,
additional gaugino and scalar masses are generated.
As we want to fully incorporate Yukawa coupling effects, we treat 
Eq.~(\ref{murayama mass}) as a boundary condition at the scale $M_Q$
and numerically evolve the masses to the weak scale across the
$S$-fields and $\phi$-fields thresholds.  
We refer the reader to
Ref.~\cite{Murayama:1997pb} for a detailed description of these
thresholds.

In the subsequent sections, 
we will investigate this LFV phenomena employing several plausible
assumptions about the mixing angles and masses of the neutrino sector.
We will find that the LFV observables
$\mu \rightarrow e \gamma$,
$\mu - e$ conversion in nuclei and $\tau \rightarrow \mu
\gamma$ can grow significantly above the SM predictions
in several interesting cases, and therefore the ideas can be tested
by present experiment and future upgrades of these LFV experiments.


\subsection{Degenerate right-handed neutrinos $M_{R_3}=M_{R_2}=M_{R_3}$}

\begin{figure}[t]
\centering
\includegraphics*[width=13cm]{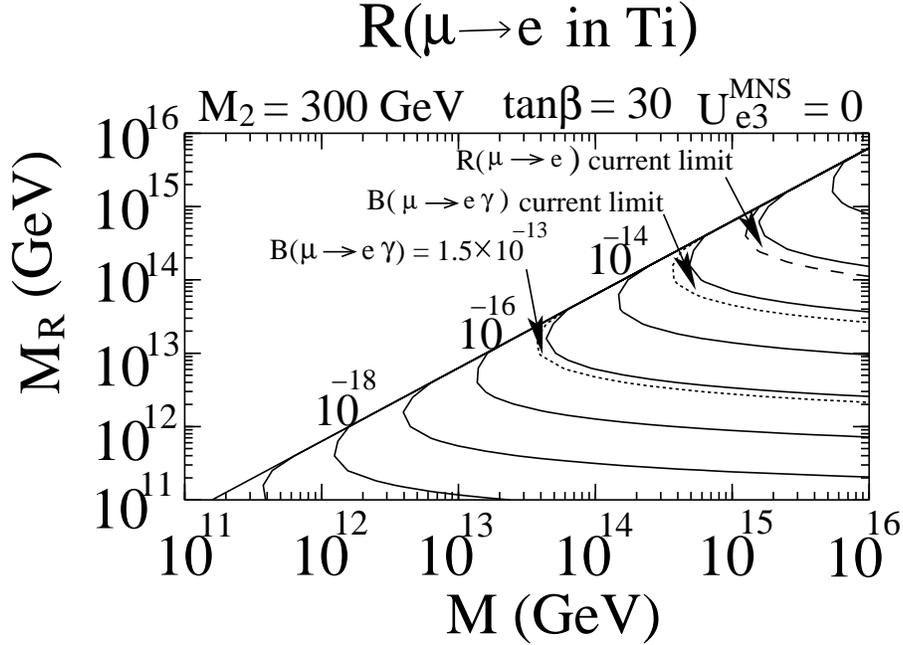}
\caption{The $\mu- e$ conversion rates
$R(\mu \rightarrow e~{\rm in~Ti})$ as a function of
the messenger scale $M$ and the degenerate 
right-handed neutrino scale $M_R$. Here we take $U_{e3}^{\rm MNS}=0$
and $\tan\beta=30$. We assume mGMSB with $n_m=1$. The low-energy wino
mass ($M_2$) is set to be 200 GeV, which fixes the entire superpartner
mass spectrum.
Current limits on $R(\mu \rightarrow e~{\rm in~Ti})$ and $B(\mu
\rightarrow e \gamma)$, and a future sensitivity at MEG [$B(\mu \rightarrow
e \gamma)=1.5\times 10^{-13}$] are also shown. Future $\mu-e$ conversion
experiments may reach a sensitivity close to 
$R(\mu -e)\simeq 10^{-18}$. Note, there is no LFV above the diagonal
line as $M_R$ is above the messenger scale.}
\label{R_mec}
\end{figure}

First we consider a case in which all right-handed neutrinos are degenerate
$M_{R_3}=M_{R_2}=M_{R_3}\equiv M_R$. 
In this case, Eq.~(\ref{yy}) is written by
\beq
(y_\nu^\dagger y_\nu)_{\mu e}\simeq U^{\rm MNS}_{\mu 3} U^{\rm MNS}_{e 3 *}
\kappa_3^D M_{R_3}+U^{\rm MNS}_{\mu 2} U^{\rm MNS}_{e 2 *} \kappa_2^D
M_{R_2}.
\label{yy_degenerate}
\eeq
Note that there is no $R$ matrix dependence in this case. Therefore
if neutrino masses, mixings and superpartner masses are fixed,
LFV event rates are expressed as a function of the messenger scale $M$
and the right-handed neutrino mass scale $M_R$.
In Fig.~\ref{R_mec}, the $\mu- e$ conversion rate
$R(\mu \rightarrow e~{\rm in~Ti})$ is shown as a function of
$M$ and $M_R$. Here we take $U_{e3}^{\rm MNS}=0$
and $\tan\beta=30$. We assume mGMSB with $n_m=1$ and wino mass ($M_2$)
to be $200$ GeV, which fixes the entire superpartner mass spectrum.
As can be seen from Fig.~\ref{R_mec}, as $M_R$ becomes larger,
the $\mu-e$ conversion rate gets larger as long as the messenger scale
$M$ is larger than, and not close to, $M_R$. As can be understood from
Eq.~(\ref{yy_degenerate}), the third generation right-handed neutrino
does not contribute to $\mu-e$ flavor violating slepton masses when
$U_{e3}^{\rm MNS}=0$. 
When $U_{e3}^{\rm MNS}=O(0.1)$, the third generation contribution to
$\mu-e$ LFV in slepton masses becomes dominant, and hence the $\mu-e$
conversion rate gets larger in general than what we present
in Fig.~\ref{R_mec}.
The effect of non-zero $U_{e3}^{\rm MNS}$ would be more important in
some case with non-degenerate right-handed neutrinos as we will discuss
later.

In the supersymmetric SM with right-handed neutrinos, the magnetic-moment-type
photon penguin diagram, which also induces the $\mu \rightarrow e \gamma$
process, is almost always dominant in the $\mu - e$ conversion process, 
and hence there is a relation between the
predicted $B(\mu \rightarrow e \gamma)$ and $R(\mu \rightarrow e~{\rm
in~Ti})$:
\beq
\frac{R(\mu \rightarrow e~{\rm in~Ti})}{B(\mu \rightarrow e \gamma)}
\simeq 5\times 10^{-3}.
\eeq
This relationship is what allows us to show in Fig.~\ref{R_mec} simultaneously
the current limits from $\mu -e$
conversion process [$R(\mu \rightarrow e~{\rm in~Ti})=6.1 \times
10^{-13}$]~\cite{talk_meco}, 
$\mu \rightarrow e \gamma$ process [$B(\mu \rightarrow e
\gamma)=1.2\times 10^{-11}$]~\cite{talk_meco}, 
and a future sensitivity at MEG experiment
[$B(\mu \rightarrow e \gamma)=1.5\times 10^{-13}$]~\cite{talk_meco,MEG}.
We also note that the expected sensitivity at the future MECO experiment 
for $\mu-e$ conversion will be $R(\mu\rightarrow
e)=10^{-16}$~\cite{talk_meco,MECO}, and further future experiment may be
able to reach a sensitivity $R(\mu\rightarrow e)=10^{-18}$ according to
studies in Ref.~\cite{future_me}.
Therefore, at present, the current $\mu \rightarrow e \gamma$ limit
already constrains the region $M_R>10^{14}$ GeV and $M>10^{15}$ GeV in 
Fig.~\ref{R_mec}. The future MEG and MECO experiments will be sensitive 
to the region around $M_R>10^{12-13}$ GeV and $M>10^{13-14}$ GeV.
It is extremely interesting to see from Fig.~\ref{R_mec} that
future measurements by experiments such as PRIME may be able to
reach the region above $M_R>10^{11}$ GeV and $M>10^{12}$ GeV.
When one considers that this encompasses the
range of allowed messenger mass 
from the leptogenesis considerations, Eq.~(\ref{M_condition}),
the result
indicates that LFV signals are expected to be measured provided
$M_R<M$.

\begin{figure}[t]
\centering
\includegraphics*[width=13cm]{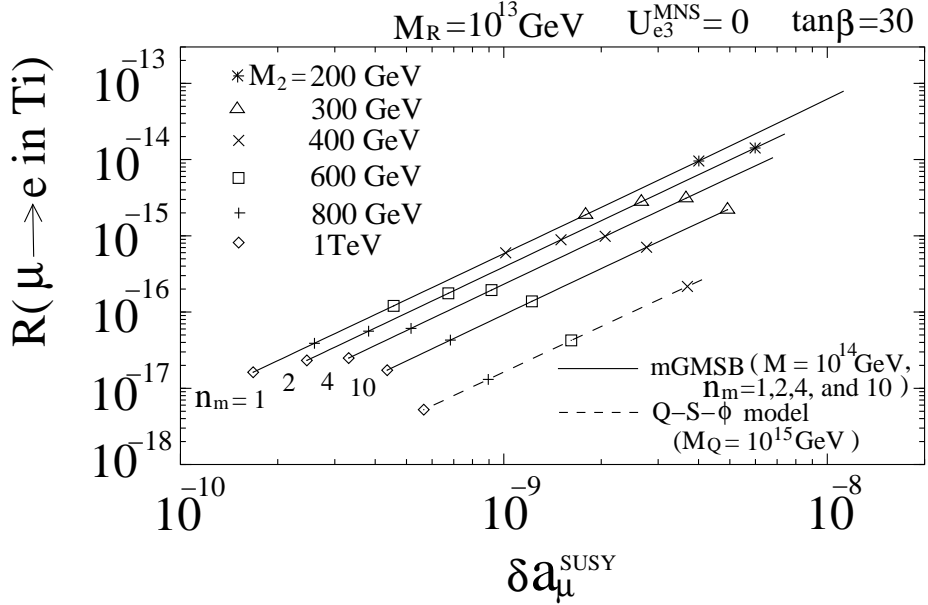}
\caption{Relations between $R(\mu \rightarrow e)$ and superpartner 
contribution
to muon $g-2$ ($\delta a_\mu^{\rm SUSY}$) as a function of low-energy
wino mass $M_2$ in mGMSB with $n_m=1,2,4,10$ and in $Q$-$S$-$\phi$
model. We assume that $M_R=10^{13}$ GeV, $U_{e3}^{\rm MNS}=0$,
$\tan\beta=30$, and the messenger scale $M=10^{14}$ GeV for mGMSB
and $M_Q=10^{15}$ GeV for $Q$-$S$-$\phi$ model. } 
\label{amu_vs_Rmec}
\end{figure}
%

In Fig.~\ref{amu_vs_Rmec}, we show relations between
$R(\mu \rightarrow e)$ and superpartner contribution to muon $g-2$
($\delta a_\mu^{\rm SUSY}$) as a function of the low-energy wino mass $M_2$
in mGMSB with $n_m=1,2,4,10$ and in the $Q$-$S$-$\phi$ model.
Here we have assumed that $M_R=10^{13}$ GeV, $U_{e3}^{\rm MNS}=0$,
$\tan\beta=30$, and the messenger scale $M=10^{14}$ GeV for mGMSB
and $M_Q=10^{15}$ GeV for $Q$-$S$-$\phi$ model.
Note that a recent updated estimate of  muon $g-2$~\cite{Davier:2003pw}
shows that the deviations of the SM prediction from the measurement 
at BNL~\cite{Bennett:2002jb} are found to be
\bea
\delta a_\mu=
\left\{
\begin{array}{l}
(22.1\pm 7.2 \pm 3.5 \pm 8.0)\times 10^{-10} ~~[e^+ e^--{\rm
based~estimate}],
\\
~(7.4\pm 5.8 \pm 3.5 \pm 8.0)\times 10^{-10} ~~[\tau-{\rm based~estimate}],
\end{array}
\right.
\eea
where the first error comes from the hadronic contribution, the second one
from the light-by-light scattering contribution and the third one from
the BNL experiment.
For fixed $M_2$, as the number of messenger multiplets $n_m$ gets
larger, $\delta a_\mu^{\rm SUSY}$ increases because slepton masses
become smaller (see Eq.~(\ref{mGMSB masses})), on the other hand, $R(\mu
\rightarrow e)$ does not 
change much because the smaller slepton masses also slightly suppress the 
LFV masses compared to the diagonal components that originated partially
from gaugino masses in the RGE running. In the $Q$-$S$-$\phi$ model,
relatively large scalar masses, which are flavor diagonal, are
additionally generated at $S$ and $\phi$ mass thresholds, which are
below the right-handed 
neutrino mass scale. Therefore the LFV masses are relatively suppressed,
and as a result, the conversion rate gets smaller. Taking into account
the allowed range of $M_Q$ ($10^{14}$ GeV $<M_Q< 10^{16}$
GeV)~\cite{Murayama:1997pb}, however, the present and future LFV
searches have great impact on sensitivity to this model too. 

LFV event rates depend on parameters in the
neutrino sector such as $M_R$ and $U_{e3}^{\rm MNS}$ as shown in
Fig.~\ref{R_mec}. On the other hand, the contribution from superpartners
to muon $g-2$ is
almost independent of these parameters because small LFV slepton masses
do not affect it. So in the parameter space in
Fig.~\ref{R_mec}, the superpartner contribution to muon $g-2$ is almost
constant, $\delta a_\mu^{\rm SUSY} \simeq 2\times 10^{-9}$.
For fixed $\delta a_\mu^{\rm SUSY}$, the messenger-model dependence
can be read from Fig.~\ref{amu_vs_Rmec}, and the dependence on the
neutrino parameters and the messenger scale can be seen from Fig.~\ref{R_mec}.
It should be noted also that $\delta a_\mu^{\rm SUSY}$ is approximately
proportional to $\tan\beta$, and $R(\mu \rightarrow e)$ is 
proportional to $\tan^2\beta$.
Since $R(\mu \rightarrow e) \propto (\delta a_\mu^{\rm
SUSY})^2$ as can be seen from Fig.~\ref{amu_vs_Rmec}~\cite{Hisano:2001qz},  
potentially the precise determination of muon $g-2$ together with
LFV searches would provide a significant handle on models with
LFV.


\subsection{Non-degenerate right-handed neutrinos 
$M_{R_3}\geq M_{R_2} \geq M_{R_1}$}

When the right-handed neutrinos are not degenerate 
$M_{R_3}\geq M_{R_2} \geq M_{R_1}$, a dependence on the $R$-matrix
enters in the slepton masses, and hence the LFV event rates
generally depend on the unknown\footnote{None of the successful
neutrino oscillation experiments can pin down
or even constrain the $R$-matrix entries. It takes additional observables,
such as LFV signals, to make progress in constraining this matrix.} 
parameters of $R_{ij}$.

First let us consider the case with $R={\bf 1}$. When $U_{e3}^{\rm MNS}=0$,
the third generation right-handed neutrino does not induce $\mu-e$
flavor violation. In this case, whatever the value
of $M_{R_3}$ is (even if $M_{R_3}>M$), the $\mu-e$ conversion
rate is the same as that found in Fig.~{\ref{R_mec}} except that now we should
read $M_R$ in Fig.~{\ref{R_mec}} as $M_{R_2}$.
However, LFV generated by the third generation  right-handed neutrino
is important in the $\tau\rightarrow \mu \gamma$ process especially
when $M>M_{R_3}$ and $M_{R_2}$ is rather small (say, less than
$10^{12}$ GeV).
%
\begin{figure}[t]
\centering
\includegraphics*[width=13cm]{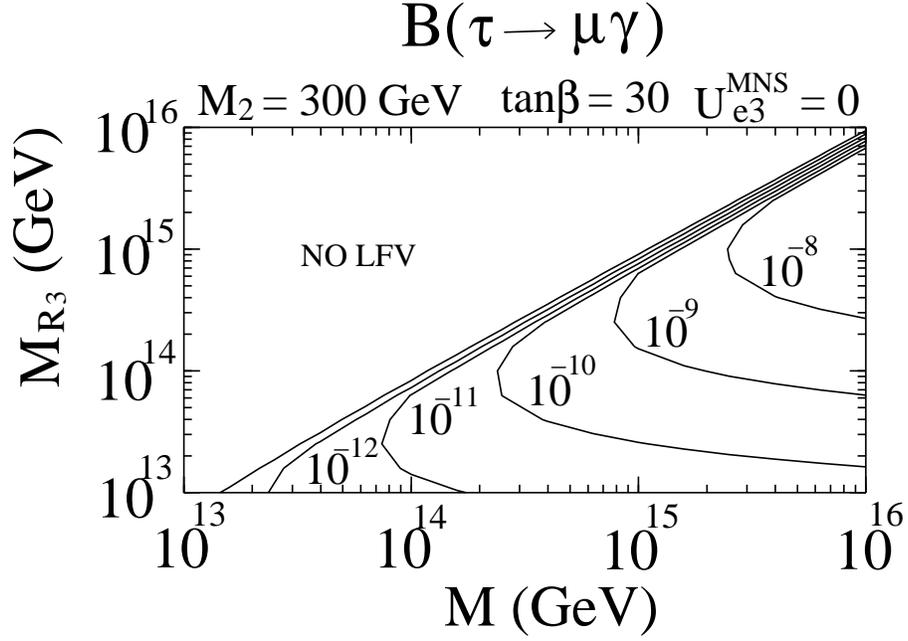}
\caption{Branching ratios of $\tau \rightarrow \mu\gamma$ process,
$B(\tau \rightarrow \mu\gamma)$ as a function of the messenger scale $M$
and the third generation right-handed neutrino mass $M_{R_3}$. 
Here we consider the hierarchical right-handed neutrinos,
$M_{R_1}=10^{10}$ GeV and $M_{R_2}=10^{12}$ GeV.
We assume mGMSB with $n_m=1$ and $M_2=300$ GeV, and 
$\tan\beta=30$, $U_{e3}^{\rm MNS}=0$ and $R={\bf 1}$. Future 
experimental sensitivity might
be as low as $10^{-9}$, in which case there is a significant
region of  $M_{R_3}$ and $M$ that this LFV observable would probe.}
\label{BR_tmg}
\end{figure}
In Fig.~\ref{BR_tmg}, branching ratio contours of $\tau \rightarrow \mu \gamma$
are shown as a function of the messenger
scale $M$ and the third generation right-handed neutrino mass $M_{R_3}$.
Here we assumed mGMSB with $n_m=1$ and $M_2=300$ GeV, $\tan\beta=30$, 
$U^{\rm MNS}_{e3}=0$ and $R={\bf 1}$. We took $M_{R_1}=10^{10}$ GeV,
$M_{R_2}=10^{12}$ GeV, 
and thus the $\mu -e$ conversion is typically small ($R(\mu \rightarrow
e)\sim 10^{-16}-10^{-17}$) because the third generation right-handed
neutrino does not contribute $(U^{\rm MNS}_{e3}=0)$ and $M_{R_2}$ is so
low.
Nevertheless, as can be seen from Fig.~\ref{BR_tmg}, the search for
$\tau\rightarrow \mu \gamma$ could be interesting and
capable of probing the region
around $M>10^{15}$ GeV and $M_{R_3}>10^{14}$ GeV if the future improved
measurement of
$B(\tau \rightarrow \mu \gamma)$ gets to the level of
$10^{-8}-10^{-9}$~\cite{Oshima:2001sh}. Therefore $\tau\rightarrow \mu
\gamma$ will be independently important in the framework of
GMSB.

Next, we consider cases where $R\neq {\bf 1}$.
We parameterize the $R$ matrix as follows:
\begin{eqnarray}
R =
\left(
\begin{tabular}{ccc}
$c_{R_2} c_{R_1}$ & $s_{R_2} c_{R_1}$ & $s_{R_1}$\\
$-s_{R_2} c_{R_3} -c_{R_2} s_{R_3} s_{R_1}$ & 
$c_{R_2} c_{R_3}-s_{R_2} s_{R_3} s_{R_1}$ & $s_{R_3} c_{R_1}$\\
$s_{R_2} s_{R_3}-c_{R_2} c_{R_3} s_{R_1}$ & 
$-c_{R_2} s_{R_3}-s_{R_2}c_{R_3} s_{R_2} $ &$c_{R_3} c_{R_1}$
\end{tabular}
\right),
\end{eqnarray}
where $s_{Ri}=\sin\theta_{R_i}$ and $c_{Ri}=\cos\theta_{R_i}$.

\begin{figure}
\centering
\includegraphics*[width=13cm]{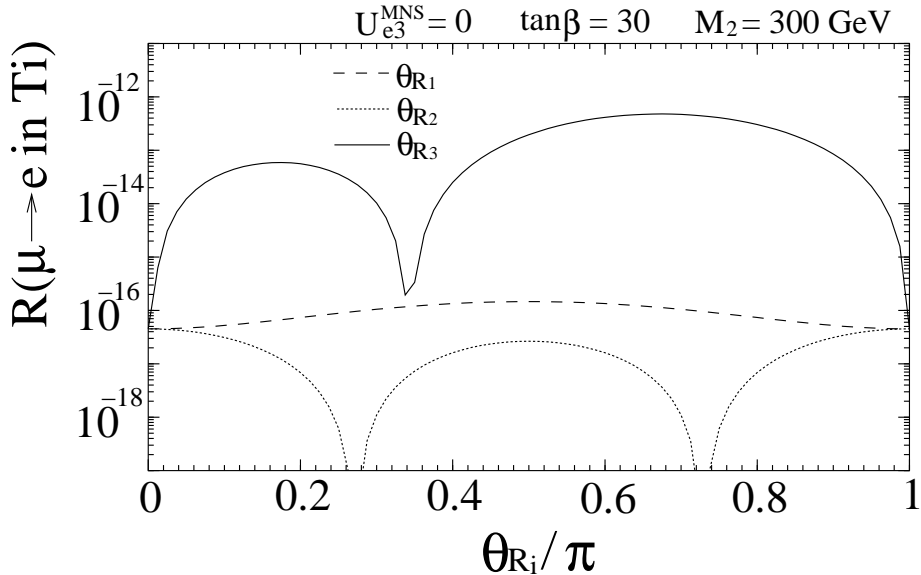}
\caption{The $\mu -e$ conversion rates, $R(\mu\rightarrow e~{\rm in
~Ti})$, as a function of an angle $\theta_{R_i}$ in the unknown $R$-matrix.
Each line in the plot is drawn by varying only one non-zero
angle $\theta_{R_i}$. We assume mGMSB with $n_m=1$ and $M_2=300$ GeV,
$\tan\beta=30$ and $U_{e3}^{\rm MNS}=0$. Here we set $M=10^{15}$ GeV, 
$M_{R_1}=10^{10}$ GeV,
$M_{R_2}=10^{12}$ GeV, and $M_{R_3}=10^{14}$ GeV.}
\label{Rmec_ve3_0}
\end{figure}

\begin{figure}
\centering
\includegraphics*[width=13cm]{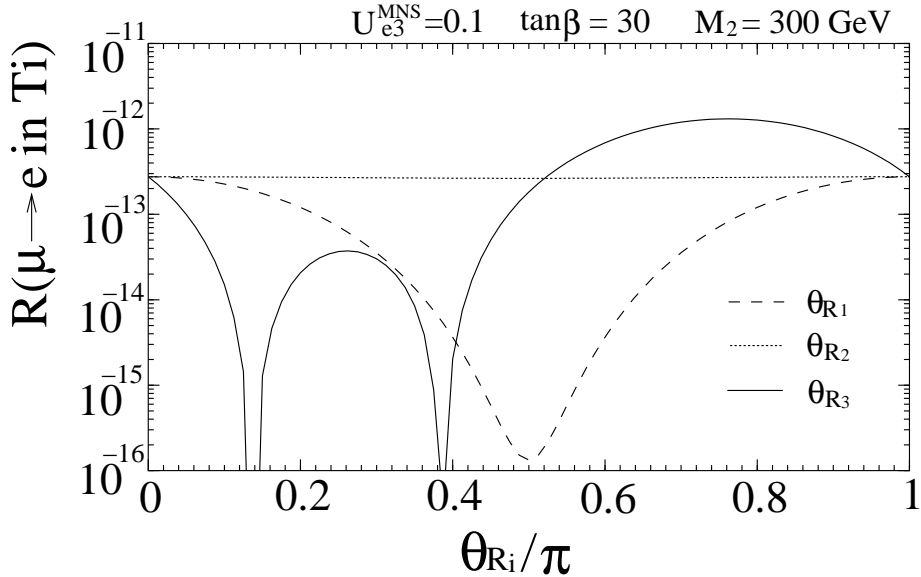}
\caption{Same as Fig.~\ref{Rmec_ve3_0} except for $U_{e3}^{\rm MNS}=0.1$.  }
\label{Rmec_ve3_01}
\end{figure}

Assuming that one of the angles $\theta_{R_i}$ is non-zero,
we present $\mu-e$ conversion rates showing the dependence on
each angle $\theta_{R_i}$ in Fig.~\ref{Rmec_ve3_0} and
\ref{Rmec_ve3_01}. We assumed mGMSB with $n_m=1$, $M_2=300$ GeV
and $\tan\beta=30$. We fixed the messenger scale and right-handed neutrino
masses to be $M=10^{15}$ GeV, $M_{R_1}=10^{10}$ GeV,
$M_{R_2}=10^{12}$ GeV, and $M_{R_3}=10^{14}$ GeV.
%
%
%
In Fig.~\ref{Rmec_ve3_0}, $U_{e3}^{\rm MNS}$ is set to zero.
As can be understood from Eq.~(\ref{yy}), due to non-zero mixing in
the $R$-matrix, several comparable terms are generated in Eq.~(\ref{yy}),
and then for some specific values of $\theta_{R_i}$,
a cancellation occurs between terms. That is why we find
some steep suppressions of $R(\mu\rightarrow e)$ in
Fig.~\ref{Rmec_ve3_0}. Otherwise the event rates in most of
Fig.~\ref{Rmec_ve3_0} are comparable or larger than those when
$R={\bf 1}$ ($\theta_{R_i}=0$).

In Fig.~\ref{Rmec_ve3_01}, we show a similar figure to Fig.~\ref{Rmec_ve3_0}
except that $U_{e3}^{\rm MNS}$ is set to be $0.1$. 
Again at some (different) values of $\theta_{R_i}$ a cancellation 
occurs and $R(\mu -e)$ dips precipitously.
However, the typical value of the event rates are often considerably
larger than those when $U_{e3}^{\rm MNS}=0$, especially
when the hierarchy between $M_{R_2}$ and $M_{R_3}$ is large.
Therefore, we conclude that we should generally expect large LFV event rates
even though the $R$-matrix is unknown.

\section{Conclusion}

We have found that if we incorporate neutrino masses and
leptogenesis into the superstructure of direct gauge-mediated supersymmetry
breaking, significant low-energy LFV signals are
possible in observables such as $\mu\to e\gamma$, $\mu-e$ conversion
in nuclei and $\tau\to \mu \gamma$. Heavy right-handed neutrinos
below the high messenger scale are what induce the large rates.
This is because the full Yukawa couplings of neutrinos are part
of the effective theory between the messenger scale and the
lower right-handed neutrino scale, thereby enabling their highly mixed
off-diagonal entries to infect the slepton masses via renormalization 
group running.  The large LFV mass entries in the slepton
mass matrix that result from the neutrino Yukawa coupling RG-effect
contribute in loops to the LFV observables.

We have shown that
small regions of parameter space are already ruled out by carefully
correlating what the theories predict for LFV observables to
what the experiments have (not) measured.  Large regions of parameter space
will be tested and probed by future experiments dedicated to significant
improvements in the search for $\mu\to e\gamma$, $\mu-e$ conversion
in nuclei and $\tau\to \mu \gamma$.  And as we have emphasized in
this article, a non-zero measurement of these LFV decays 
is not only consistent with some versions of gauge mediation,
it is to be expected.

\section*{Acknowledgements}
This work was supported in part by the Department of Energy and the
Alfred P. Sloan Foundation.

\end{document}